\documentclass[11pt]{article}
\usepackage[margin=.7in]{geometry}
\usepackage[font=small,labelfont=bf]{caption}
\usepackage{authblk}
\usepackage[utf8]{inputenc}
\usepackage{amsmath}
\usepackage{amssymb}
\usepackage{mathtools}
\usepackage{float}
\usepackage{placeins}
\usepackage{braket}
\usepackage{tikz}
 
\usepackage[
    backend=biber,
    citestyle=numeric,
    bibstyle=numeric,
    sorting=none
]{biblatex}
\usepackage{pgfplots}
\pgfplotsset{compat=newest}
\usepackage{xfrac}
\usetikzlibrary{decorations.pathreplacing}
\usetikzlibrary{positioning, backgrounds}
\usetikzlibrary{calc}
\usetikzlibrary{arrows.meta}
\usepackage{multirow}

\DeclarePairedDelimiter{\abs}{\lvert}{\rvert}%
\DeclarePairedDelimiter{\norm}{\lVert}{\rVert}%

\makeatletter
\let\oldabs\abs
\def\abs{\@ifstar{\oldabs}{\oldabs*}}
\let\oldnorm\norm
\def\norm{\@ifstar{\oldnorm}{\oldnorm*}}
\makeatother
\begin{document}
\title{Statistical Approach to Quantum Phase Estimation}

\author[1]{Alexandria J.~Moore\thanks{moore428@purdue.edu}}
\author[2]{Yuchen Wang}
\author[2]{Zixuan Hu}
\author[2]{Sabre Kais\thanks{kais@purdue.edu}}
\author[1]{Andrew M.~Weiner\thanks{amw@purdue.edu}}

\affil[1]{%
School of Electrical and Computer Engineering and Purdue Quantum Science and Engineering Institute, Purdue University, West Lafayette, Indiana 47907, USA}
\affil[2]{%
Department of Chemistry, Department of Physics and Purdue Quantum Science and Engineering Institute, Purdue University, West Lafayette, Indiana 47907, USA}

\date{April 20, 2021}
\maketitle
\begin{abstract}
We introduce a new statistical and variational approach to the phase estimation algorithm (PEA). Unlike the traditional and iterative PEAs which return only an eigenphase estimate, the proposed method can determine any unknown eigenstate-eigenphase pair from a given unitary matrix utilizing a simplified version of the hardware intended for the Iterative PEA (IPEA). This is achieved by treating the probabilistic output of an IPEA-like circuit as an eigenstate-eigenphase proximity metric, using this metric to estimate the proximity of the input state and input phase to the nearest eigenstate-eigenphase pair and approaching this pair via a variational process on the input state and phase. This method may search over the entire computational space, or can efficiently search for eigenphases (eigenstates) within some specified range (directions), allowing those with some prior knowledge of their system to search for particular solutions. We show the simulation results of the method with the Qiskit package on the IBM Q platform and on a local computer.
\end{abstract}

\section{Introduction}
Efficient spectral decomposition of large matrices is a key component to many optimization and machine learning algorithms, with applications ranging from factoring and searching algorithms to computational chemistry~\cite{kais2014quantum}. On classical computers, spectral decomposition scales super-linearly with the system dimension~\cite{Liberty20167}, making it intractable for large problems. Due to the utility of spectral decomposition and its classical limitations, quantum approaches to spectral decomposition and eigenvalue estimation have been pursued~\cite{daskin2011decomposition}. One significant approach is the quantum phase estimation algorithm (PEA)~\cite{nielsenChuang} -- a means of determining unknown eigenphases of a unitary matrix -- which is a key subroutine in a number of quantum algorithms including Shor's factoring algorithm~\cite{shor1994algorithms}, quantum principal component analysis~\cite{llyod2014}, the generalized Grover's search algorithm~\cite{generalizedgrover2018}, and quantum simulations~\cite{wang2008quantum,aspuru2005simulated,daskin2018direct}.

Near-term quantum systems operate in the noisy intermediate-scale quantum (NISQ) regime~\cite{preskill2018quantum}, facing restrictions on both circuit depth and breadth due to decoherence and gate infidelity. Consequently, interest in the traditional PEA \cite{nielsenChuang} and quantum principal component analysis \cite{llyod2014} has been channeled toward developments in the \textit{iterative} PEA (IPEA) \cite{dobvsivcek2007} -- a method which estimates an unknown phase over multiple circuit iterations --  allowing for significant reduction in both qubit usage (circuit breadth) and controlled-gate operations (circuit depth). The IPEA has been demonstrated on photonic systems \cite{paesani2017}. On the other hand, variational quantum algorithms (VQA) -- which use a classical computer to control and optimize the parameters applied in a quantum circuit -- have been developed for a variety of problems as they leverage the speedup of quantum algorithm with lower-depth circuits \cite{wang2019accelerated,peruzzo2014variational}. 

Here, we introduce a quantum-classical hybrid algorithm combining the PEA with the VQA -- which we call the \textit{Statistical} PEA (SPEA) -- and show preliminary simulation results on the IBM Q platform with the Qiskit package~\cite{Qiskit} as well as simulations on a local computer. The method is able to determine any unknown eigenstate-eigenphase pair from a unitary matrix by utilizing hardware intended for the IPEA. Further, the SPEA can be applied repeatedly to obtain a full spectral decomposition. The SPEA may be compared to other variational quantum eigensolvers \cite{purityVariationalDiag2019, VQEColes2020,o2016scalable}, the primary difference being other variational eigensolvers work directly on a (Hermitian)  matrix encoded as a quantum state using specially designed quantum circuits. The SPEA assumes access to a gate representation of the unitary exponentiation of the state -- or assumes simultaneous availability of several copies of the quantum state to approximate the quantum gate à la \cite{llyod2014}. In return, the SPEA requires a polynomially-reduced number of (classical) optimization parameters -- as it optimizes for a single eigenstate, rather than diagonalize the entire matrix simultaneously -- and directly delivers eigenstate-eigenphase pairs (whereas other approaches may allow on-demand generation of eigenstates, but require tomography if knowledge of the state is needed). The SPEA is also able to search for eigenphases within specified ranges, allowing those with some prior knowledge of their system to search for particular solutions, whether ground state (near minimum eigenphase), principle (near-maximal eigenphase), or any other region of interest.

This paper is organized as follows: Section~\ref{Sec:PEA intro} reviews the traditional and iterative PEA and introduces a statistical metric  $\mathcal{C}$ for quantifying the proximity of any given input-state to its closest eigenstate. Section~\ref{Sec:SPEA theory} describes the Statistical PEA and discusses the connections between the $\mathcal{C}$ factor and the quality (in terms of proximity) of the derived eigenstate-eigenphase pairs (with the derivation details in Appendix~\ref{appendix:CMetric}). Section~\ref{Sec:SPEA theory} also outlines the optimization process for obtaining the eigenstate-eigenphase pairs. Simulation results on different platforms are reported and discussed in Section~\ref{Sec:SPEA simulation}; methodology details are provided in  Appendix~\ref{appendix:IBMQ} and~\ref{appendix:eigendecomp}. We conclude with a discussion on the performance of the SPEA and propose future directions and applications of the method in Section~\ref{Sec:conclusion}.

\section{Phase Estimation Algorithms}
\label{Sec:PEA intro}
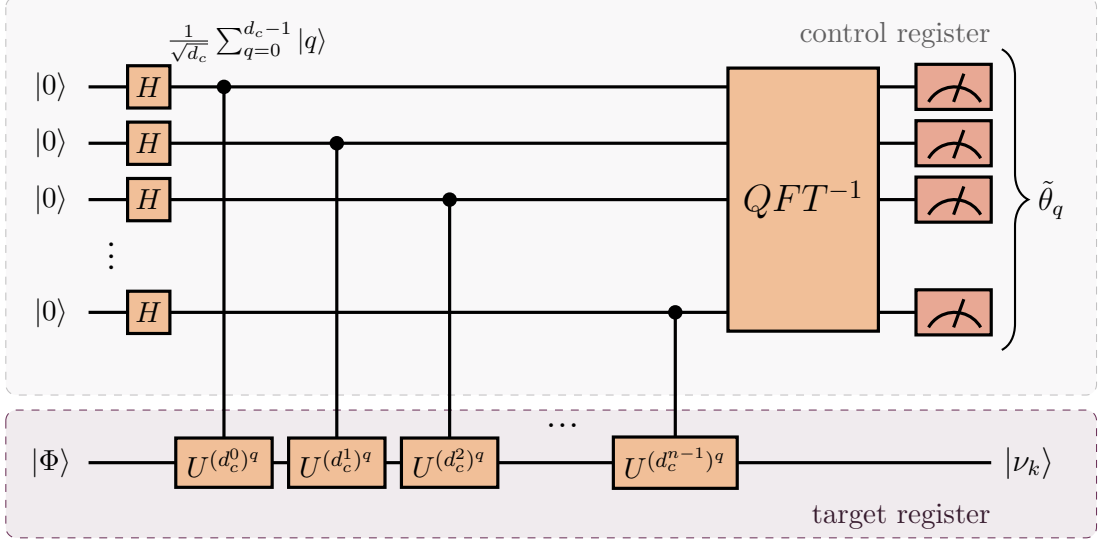
\begin{figure}
    \centering
    \definecolor{controlColor}{HTML}{C2C1C2}	
\definecolor{targetColor}{HTML}{6E355B}	
\definecolor{gateColor}{HTML}{F1BF98}	
\definecolor{measureColor}{rgb}{.909,.66,.58}	

\begin{tikzpicture}
\newcommand\Yt{-6}
\newcommand\delc{2}
\newcommand\johnlist{{"$U^{(d_c^0)^q}$","$U^{(d_c^1)^q}$","$U^{(d_c^2)^q}$","$U^{(d_c^3)^q}$","$U^{(d_c^{n-1})^q}$"}}
\def\measure at (#1,#2){
    \draw[black, very thick, fill=measureColor] (#1-.5,#2+.3) rectangle ++(1,-.6);
    \draw[very thick] (#1+.36,#2-.2) arc (30:150:.4);
    \draw[very thick] (#1,#2-.2) -- ++(.136,.375);
}
\def\Yarray{{-1, -1.75, -2.5, -3.25, -4}}

\draw[draw=controlColor, dashed, rounded corners, fill opacity=0.1, fill=controlColor] (-.6,\Yarray[0]+1.2) rectangle (13.9,\Yarray[4]-1.1);
\node[minimum size=16pt, inner sep=0pt, anchor=south east, align=right, text = controlColor!50!black] at (12.5,-.6) {control register};

\draw[draw=targetColor, dashed, rounded corners, fill opacity=0.1, fill=targetColor] (-.6,\Yarray[4]-1.3) rectangle (13.9,\Yarray[4]-3);
\node[minimum size=16pt, inner sep=0pt, anchor=south east, align=right, text = targetColor!50!black] at (12.5,\Yarray[4]-3) {target register};

    \node[circle, minimum size=16pt, inner sep=0pt] at (0,\Yt) {$\ket{\Phi}$};
    \draw[black, very thick] (.5,\Yt) -- (12.5,\Yt);
    \node[circle, minimum size=16pt, inner sep=0pt] at (13,\Yt) {\large $\ket{\nu_k}$};

\foreach \x in {0,1,2,4}{
    \node[circle, minimum size=16pt, inner sep=0pt] at (0,\Yarray[\x]) {$\ket{0}$};
    \draw[black, very thick]
        (.5,\Yarray[\x]) --
        +(.8, 0) node[rectangle, draw=black, fill=gateColor, inner sep=0pt, minimum size=16pt] (name) {$H$} -- 
        +(12,0);
    \measure at (12,\Yarray[\x]);
    \draw[black, very thick]
        (.8+1.5+1.5*\x, \Yarray[\x]) node[circle,inner sep=0pt,draw, fill=black, minimum size=.15cm] {.} --
        +(0, \Yt-\Yarray[\x]) node[rectangle, draw=black, fill=gateColor, inner sep=4pt, minimum size=16pt] (name) {\pgfmathparse{\johnlist[\x]}\pgfmathresult};
    
}
\draw[black, very thick, fill=gateColor] (9,-.75) rectangle (11,-4.25) node[pos=.5] {\Large $QFT^{-1}$};

\node[circle, rotate=90, minimum size=24pt] at (.8,\Yarray[3]) {\LARGE ...} ;
\node[circle, minimum size=24pt] at (.8+1.5+1.5*3,-5.5) {\LARGE ...} ;

\draw [thick, decorate,decoration={brace,amplitude=10pt,raise=4pt},yshift=0pt]
(12.5,\Yarray[0]+.5) -- (12.5,\Yarray[4]-.5) node [black,midway,xshift=0.8cm] {\large
$\tilde{\theta}_q$};

\node[circle, minimum size=16pt, inner sep=0pt, anchor=west] at (1.5,\Yarray[0]+.6) {\small $\frac{1}{\sqrt{d_c}} \sum_{q=0}^{d_c-1}\ket{q}$};

\end{tikzpicture}
    \caption{The traditional PEA using $n$ control $d_c$-dimensional qudits in the control register. The quantum gates are colored in orange and the measurement gates in red. The target register is highlighted in purple and control register in grey. Note that in the $d_c >2$ case, the $H$ ``Hadamard'' gate acts as a discrete quantum Fourier transform (QFT) gate. Additionally, in the $d_c >2$ case the control gates acts as MVCGs, applying the gate to the $q^\text{th}$ power when the control qudit is in state $\ket{q}$. The circuit will estimate the phase $\tilde{\theta}_k$ of eigenstate $\ket{\nu_k}$ to precision $d_c^{-(n)}$. The estimate of eigenphase $\theta_k$ is returned with probability $\lvert \braket{\Phi | \nu_k }\rvert^2$.}
    \label{fig:tradPEA}
\end{figure}

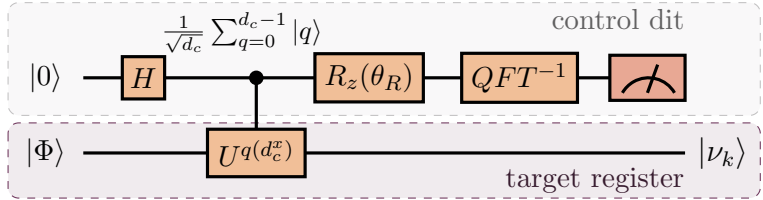
\begin{figure}
    \centering
    \definecolor{controlColor}{HTML}{C2C1C2}	
\definecolor{targetColor}{HTML}{6E355B}	
\definecolor{gateColor}{HTML}{F1BF98}	
\definecolor{measureColor}{rgb}{.909,.66,.58}	
\begin{tikzpicture}
\newcommand\Yt{-2}
\newcommand\delc{2}
\def\measure at (#1,#2){
    \draw[black, very thick, fill=measureColor] (#1-.5,#2+.3) rectangle ++(1,-.6);
    \draw[very thick] (#1+.36,#2-.2) arc (30:150:.4);
    \draw[very thick] (#1,#2-.2) -- ++(.136,.375);
}
\def\Yarray{{-1, -1.75, -2.5, -3.25, -4}}

\draw[draw=controlColor, dashed, rounded corners, fill opacity=0.1, fill=controlColor] (-.5,\Yarray[0]+1) rectangle (9.5,\Yarray[0]-.5);
\node[minimum size=16pt, inner sep=0pt, anchor=south east, align=right, text = controlColor!50!black] at (8.5,\Yarray[0]+.5) {control dit};
\draw[draw=targetColor, dashed, rounded corners, fill opacity=0.1, fill=targetColor] (-.5,\Yarray[0]-.6) rectangle (9.5,\Yarray[0]-1.6);
\node[minimum size=16pt, inner sep=0pt, anchor=south east, align=right, text = targetColor!50!black] at (8.5,\Yarray[0]-1.65) {target register};

    \node[circle, minimum size=16pt, inner sep=0pt] at (0,\Yt) {$\ket{\Phi}$};
    \draw[black, very thick] (.5,\Yt) -- (8.5,\Yt);
    \node[circle, minimum size=16pt, inner sep=0pt] at (9,\Yt) {\large $\ket{\nu_k}$};

    \node[circle, minimum size=16pt, inner sep=0pt] at (0,-1) {$\ket{0}$};
    \draw[black, very thick]
        (.5,-1) --
        ++(.8, 0) node[rectangle, draw=black, fill=gateColor, inner sep=0pt, minimum size=16pt] (name1) {$H$} -- 
        ++(3,0) node[rectangle, draw=black, fill=gateColor] (name2) {$R_z(\theta_R)$} --
        ++(2,0) node[rectangle, draw=black, fill=gateColor] (name3) {$QFT^{-1}$} --
        ++(2,0) node (controlEnd) {};
    \measure at (8,-1);
    \draw[black, very thick]
        (.8+2, -1) node[circle,inner sep=0pt,draw, fill=black, minimum size=.15cm] {.} --
        +(0, \Yt+1) node[rectangle, draw=black, fill=gateColor, inner sep=4pt, minimum size=16pt] (name) {$U^{q(d_c^{x})}$};

\node[circle, minimum size=16pt, inner sep=0pt, anchor=west] at (1.5,\Yarray[0]+.6) {\small $\frac{1}{\sqrt{d_c}} \sum_{q=0}^{d_c-1}\ket{q}$};

\end{tikzpicture}
    \caption{The iterative PEA. See Fig.~\ref{fig:tradPEA} for color-conventions and notes on the MVCG. To retrieve $n$ dits of the eigenphase (i.e. phase precision $\pm 1/d_c^{n}$), run the circuit with $x = n-1$ and $\theta_R=0$; the measured control state is the $n^{\text{th}}$ base-$d_c$ dit of the unknown phase. Proceed to run the circuit for $x=n-2$ and set $\theta_R$ according to the previous control results; the measured control state is the $(n-1)^{\text{th}}$ dit. Continue the process iteratively until $x=0$ and the entire phase is recovered. Note that the iterative method is diagrammed for a single qudit control, but may be realized with any number of control qudits, similar to the traditional PEA}
    \label{fig:iterativePEA}
\end{figure}

\begin{figure}
    \centering
    \definecolor{cm2}{rgb}{.0313,.3607,.0666}	
\definecolor{cm3}{rgb}{.8,.1215,.21} 	
\definecolor{cm4}{rgb}{.84,.7451,.0431}	
\definecolor{cm8}{rgb}{.3961,.1215,.7484}	
\definecolor{cm16}{rgb}{.7058,.3674,.0706}	

\pgfplotsset{
every tick label/.append style = {font=\tiny},
every axis label/.append style = {font=\scriptsize},
    compat=1.12,
    /pgf/declare function={
    	f2(\x) = 2*cos(deg(\x)*1);
    	f3(\x) = f2(\x)  + 2*cos(deg(\x)*2);
    	f4(\x) = f3(\x)  + 2*cos(deg(\x)*3);
    	f5(\x) = f4(\x)  + 2*cos(deg(\x)*4);
    	f6(\x) = f5(\x)  + 2*cos(deg(\x)*5);
    	h(\x,\k,\dc) = 2*((\dc)-(\k))*cos(deg(\x)*(\k));
    	g2(\x) = 1/2 + h(\x,1,2)/4;
    	g3(\x) = 1/3 + (h(\x,1,3)+h(\x,2,3))/9;
    	g4(\x) = 1/4 + (h(\x,1,4)+h(\x,2,4)+h(\x,3,4))/16;
    	g8(\x) = 1/8 + (h(\x,1,8)+h(\x,2,8)+h(\x,3,8)+h(\x,4,8)+h(\x,5,8)+h(\x,6,8)+h(\x,7,8))/64;
    	g16(\x) = 1/16 + (h(\x,1,16)+h(\x,2,16)+h(\x,3,16)+h(\x,4,16)+h(\x,5,16)+h(\x,6,16)+h(\x,7,16)+h(\x,8,16)+h(\x,9,16)+h(\x,10,16)+h(\x,11,16)+h(\x,12,16)+h(\x,13,16)+h(\x,14,16)+h(\x,15,16))/256;
    },
}
\begin{tikzpicture}
	\begin{axis}[
		name = MyAxis,
		no markers,
		domain=-3.14:3.14,
		samples=250,
		thick,
		reverse legend,
		ymin=0, ymax=1.05,
		xmin=-3.15, xmax=3.15,
		axis x line=center, 
		axis y line=left, 
		xtick={
    		-3.1416,   -2.3562,   -1.5708,   -0.7854,   0,    0.7854,    1.5708,    2.3562,    3.1416
		},
		xticklabels={
    		$\sfrac{-1}{2}$, , $\sfrac{-1}{4}$,  , $0$,  ,  $\sfrac{1}{4}$, , $\sfrac{1}{2}$
		},
    	ytick={
    		0,    0.0625,    0.1250,    0.1875,    0.2500,    0.3125,    0.3750,    0.4375,    0.5000,    0.5625,    0.6250,    0.6875,    0.7500,    0.8125,    0.8750,    0.9375,    1.0000
		},
		yticklabels={
    		$0$, , $\sfrac{1}{8}$, , $\sfrac{1}{4}$, , $\sfrac{3}{8}$,  ,  $\sfrac{1}{2}$,  , $\sfrac{5}{8}$, , $\sfrac{3}{4}$, , $\sfrac{7}{8}$, , $1$
},
		grid = major,
		xlabel = {\normalsize Phase ($\Delta\theta$)},
		ylabel = {\normalsize Probability, $\mathcal{P}_0(\Delta\theta)$},
		xlabel style = {at = {(ticklabel cs:.5)}, anchor=south, yshift=-2ex, },
		ylabel style = {at = {(ticklabel cs:.5)}, anchor=center, xshift=-0ex, yshift=1.25ex, rotate=0},
		ticklabel style = {font=\small},
		legend style={font=\tiny},
]
        \addplot+ [cm16]{g16(x)};
        \addlegendentry{${d_c=16}$};
        \addplot+ [cm8]{g8(x)};
        \addlegendentry{${d_c=8}$};
        \addplot+ [cm4]{g4(x)};
        \addlegendentry{${d_c=4}$};
        \addplot+ [cm3]{g3(x)};
        \addlegendentry{${d_c=3}$};
        \addplot+ [cm2] {g2(x)};
        \addlegendentry{${d_c=2}$};
        
        \addplot[forget plot, only marks, mark=diamond, mark options={scale=.6}, fill=cm16, draw=cm16] coordinates {(.3186, 0.048453) (-.3186, 0.048453)};
        \addplot[forget plot, only marks, mark=diamond, mark options={scale=.6}, fill=cm8, draw=cm8] coordinates {(0.63335, 0.052513) (-0.63335, 0.052513)};
        \addplot[forget plot, only marks, mark=diamond, mark options={scale=.6}, fill=cm4, draw=cm4] coordinates {(1.2309, 0.074074) (-1.2309, 0.074074)};
        \addplot[forget plot, only marks, mark=diamond, mark options={scale=.6}, fill=cm3, draw=cm3] coordinates {(1.5708, 0.11111) (-1.5708, 0.11111)};
   \end{axis}
\end{tikzpicture}
    \caption{Plot of $\mathcal{P}_0(\Delta\theta)$ (Equation~\ref{Eq:Prob0}) from $\Delta\theta\in[-.5,.5)$ for various $d_c$. Note that $\mathcal{P}_0$ has infinite domain with period $1$. Probability of the ($d_c$-level) control register of an iterative PEA collapsing to $\ket{0}$ as a function of difference between the eigenphase $\theta$ and the applied rotation $\theta_R$, $\Delta\theta \equiv \theta - \theta_R$ for an eigenstate input. Note that when the applied rotation matches the eigenphase ($\Delta\theta =0$), the control collapses to $\ket{0}$ deterministically. Denote the region around $\Delta\theta =0$ (from dot to dot) as the central lobe of $\mathcal{P}_0(\Delta\theta)$, and the small lobes with local maxima outside of it as the sidelobes. See that the higher the system's dimensionality, the narrower the probability curve's central lobe and the lower local maxima in the sidelobes. Note that $d_c=2$ has no sidelobes (the probability is monotonic on either side of the central lobe). Also note $\mathcal{P}(\Delta\theta)=0$ for $\Delta\theta = d_c^{-1}$ and the width of the central lobe is therefore $\Delta\theta_{\text{FWFM}} = 2d_c^{-1}$.}
    \label{fig:Pcurves}
\end{figure}
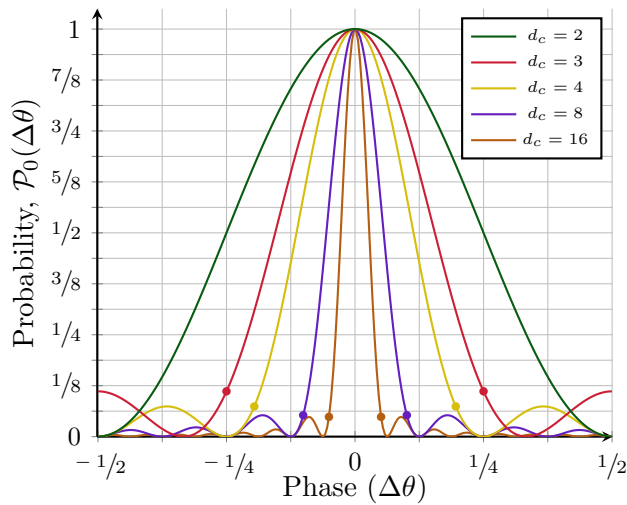

Traditional PEA implementations, diagrammed in Figure~\ref{fig:tradPEA}, take any given unitary $\hat{U}$ and any given eigenstate $\ket{\nu}$ of $\hat{U}$ and return the corresponding eigenphase $\theta$ where
\begin{align} \label{eq:eigen}
    \hat{U}\ket{\nu} = e^{i2\pi\theta} \ket{\nu}.
\end{align}
The (approximate) eigenphase $\tilde{\theta}\in [0,1)$ (equivalently, $\in [-.5,.5)$) may be directly measured on the control qubits (or qudits, when the control is $d_c$-dimensional) of the PEA. The target register is typically unmeasured during the process. For an arbitrary target register input $\ket{\Phi}$, the probability of the circuit representing a particular eigenstate $\ket{\nu_{k}}$ and the associated eigenphase $\theta_{k}$ is $\abs{\braket{ \nu_k|\Phi }}^2$. If $\ket{\Phi}$ is not itself an eigenstate, the eigenphase retrieved varies each time the PEA circuit is run. The prototypical PEA thus approximates a particular $\theta$ in a single trial.

The traditional PEA requires large quantum circuits which are often unreliable in the NISQ regime. To overcome hardware constraints, the iterative PEA (IPEA) was developed. The IPEA significantly reduces circuit depth requirements by approximating a particular $\theta$ one qubit (or $d_c$-level qudit) at a time, starting from the least significant qubit (qudit). The IPEA requires a rotation gate -- a linear phase across the control register -- to ``subtract'' off eigenphase information determined in previous iterations. (I.e. if the quantum circuit's state before $R_z(\theta_R)$ is $\sum_q\alpha_q\ket{q}\ket{\Phi_q}$, then after the rotation gate the quantum circuit's state is $\sum_q\alpha_q e^{-iq2\pi \theta_R}\ket{q}\ket{\Phi_q}$.) The iterative PEA, as the name suggests, requires a number of iterations equal to the number of bits (dits) of precision desired from the eigenphase. Additionally, the input to the target register of an IPEA must either be an eigenstate (and identically prepared each iteration) or the previous iteration's output must propagate forward and serve as the next iteration's input.

An IPEA circuit is diagrammed in Figure~\ref{fig:iterativePEA}. In the general case, the control qudit may be high-dimensional ($d_c-$level). In this case, the Hadamard gates represent a $d_c$-dimensional quantum Fourier transform gate and the control-$\hat{U}$ gate is a multi-level control gate (MLCG)~\cite{lu2020quantum}: when the control state is $\ket{q}$, a $\hat{U}^q$ gate is applied to the target register. Consider the IPEA in its ``last'' iteration's settings ($x=0$ in Figure~\ref{fig:iterativePEA}). When the target register is an eigenstate $\ket{\Phi} = \ket{\nu}$ and the rotation gate is used to subtract off phase $2\pi\theta_R = 2\pi\theta$, the control dits deterministically collapse to state $\ket{0}$. When either the target input is not an eigenstate and/or $\theta_R$ is not the corresponding eigenphase, the control dits will collapse to $\ket{0}$ with non-unity probability.

Indeed, for eigenstate input $\ket{\nu}$ with a $d_c$-level control, the final state of the control qudit before measurement is
\begin{align}
\ket{\Psi_C} = \frac{1}{d_c} \sum_{q=0}^{d_c-1} \sum_{n=0}^{d_c-1} e^{2\pi i n (\theta - \theta_R - \frac{q}{d_c})}\ket{q}
\end{align}
where $\theta$ is the eigenphase of $\ket{\nu}$ and $-\theta_R$ ($\theta_R\in [0,1)$) is the rotation applied by the rotation gate. The probability of measuring the system in output bin $\ket{0}$ is 
\begin{align}\label{Eq:Prob0}
\begin{split}
\mathcal{P}_{\theta}(-\theta_R) &=  \abs{\braket{0 | \Psi_C}}^2 = \frac{1}{d_c^2} \abs{\sum_{n=0}^{d_c-1} e^{2\pi i n (\theta - \theta_R)}}^2 \\
&=  \mathcal{P}_0(\theta - \theta_R).
\end{split}
\end{align}
$\mathcal{P}_{\theta}(-\theta_R)$ goes to one as $\theta_R$ approaches $\theta$, as shown in Figure~\ref{fig:Pcurves}. In the most general case, where the target register is an arbitrary (non-eigenstate) input state $\ket{\Phi}$ and the rotation gate subtracts off phase $2\pi\theta_R$, the probability that the control qudits will collapse to to $\ket{0}$ is

\begin{align}\label{eq:Chat}
\begin{split}
\mathcal{C}(\ket{\Phi}, \theta_R) &= \sum_{k=0}^{d_t - 1} \abs{\braket{\nu_k | \Phi}}^2 \mathcal{P}_{\theta_k}(-\theta_R) \\
    &= \frac{1}{d_c^2} \sum_{k=0}^{d_t - 1} \abs{\braket{\nu_k | \Phi}}^2 \abs{\sum_{n=0}^{d_c-1} e^{2\pi i n (\theta_k - \theta_R)}}^2
\end{split}
\end{align}

Where $\hat{U}$ is $d_t$-by-$d_t$-dimensional and the target register is $d_t$-dimensional. Appreciate that $\mathcal{C}(\ket{\Phi}, \theta_R) = 1$ if and only if $\ket{\Phi}$ is an eigenstate and $\theta_R$ is its corresponding eigenphase.

\FloatBarrier
\section{Statistical Approach to PEAs}
\label{Sec:SPEA theory}
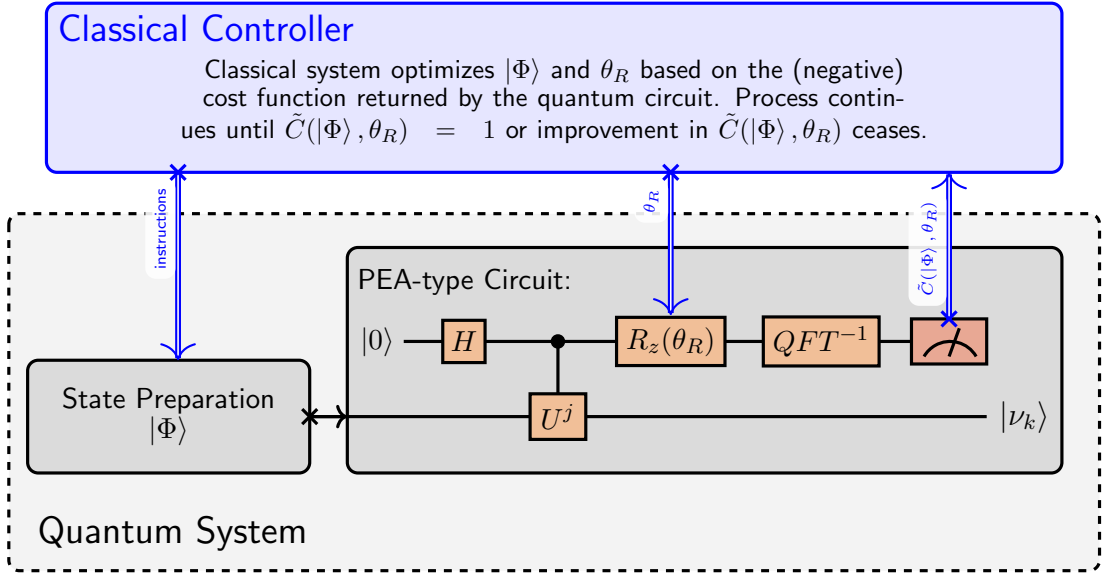
\begin{figure}
    \centering
    \definecolor{controlColor}{HTML}{C2C1C2}	
\definecolor{targetColor}{HTML}{6E355B}	
\definecolor{gateColor}{HTML}{F1BF98}	
\definecolor{measureColor}{rgb}{.909,.66,.58}	

\begin{tikzpicture}[font=\sffamily]
\tikzset{classInfo/.style={blue, double, thick}};
\tikzset{quantInfo/.style={black, very thick}};
\tikzset{classicTextNode/.style={rotate=90, left=.02, anchor=south, fill=white, opacity=.9, text opacity=1, rectangle, rounded corners}};

\newcommand\Yt{-2}
\newcommand\delc{2}
\def\measure at (#1,#2){
    \draw[black, very thick, fill=measureColor] (#1-.5,#2+.3) rectangle ++(1,-.6);
    \draw[very thick] (#1+.36,#2-.2) arc (30:150:.4);
    \draw[very thick] (#1,#2-.2) -- ++(.136,.375);
}
\def\Yarray{{-1, -1.75, -2.5, -3.25, -4}}

\newcommand\xWidth{.1}
\def\xTip[#1] at (#2,#3){
	\draw[very thick, #1] (#2-\xWidth, #3+\xWidth) -- (#2+\xWidth, #3-\xWidth);
	\draw[very thick, #1] (#2-\xWidth, #3-\xWidth) -- (#2+\xWidth, #3+\xWidth);
}

\newcommand\classY{2.5};
\newcommand\delClassYY{-2.25};
\newcommand\delXX{14.5};
\newcommand\delQuantYY{-4.75};
\newcommand\peaXoffset{-1.0};
\newcommand\peaYoffset{-1};
\coordinate (oX) at (-5.5,0);
\coordinate (delXX) at (\delXX,0);
\coordinate (classY) at (0,\classY);
\coordinate (quantY) at (0,-.3);
\coordinate (delClassYY) at (0,\delClassYY);
\coordinate (delQuantYY) at (0,\delQuantYY);

\draw[blue, very thick, rounded corners, fill=blue, fill opacity=.1, text opacity=1] ($(oX) + (classY) + (.5,0)$) node[anchor=north west] {\Large Classical Controller}rectangle ++($(delXX) + (delClassYY) + (-1,0)$);
\node[text width=14cm, anchor=north, align=center] at ($(oX) + (classY) + (\delXX/2,-.6)$) {\small \baselineskip=2pt Classical system optimizes $\ket{\Phi}$ and $\theta_R$ based on the (negative) cost function returned by the quantum circuit. Process continues until $\tilde{C}(\ket{\Phi}, \theta_R)=1$ or improvement in $\tilde{C}(\ket{\Phi}, \theta_R)$ ceases.\par};
\draw[black, very thick, dashed, rounded corners, fill=black, fill opacity=.05, text opacity=1] ($(oX) + (quantY)$) rectangle ++($(delXX) + (delQuantYY)$);
\node[rectangle, align=left, anchor=west] at ($(oX) + (quantY) + (.25, \delQuantYY+.5)$) {\Large Quantum System};

\draw[black, very thick, rounded corners, fill=black, fill opacity=.1, text opacity=1] (\peaXoffset-4.25,\peaYoffset+\Yt+.75) rectangle ++(3.75,-1.5) node[pos=.5, rectangle, minimum size=16pt, inner sep=0pt, align=center, text width=4cm] {\baselineskip=6pt State Preparation \par $\ket{\Phi}$ \par};

\draw[black, very thick, rounded corners, fill=black, fill opacity=.1, text opacity=1] (\peaXoffset0,\peaYoffset+.25) rectangle (\peaXoffset+9.5,\peaYoffset+\Yt-.75);
\node[rectangle, align=left, right] at  (\peaXoffset-0, \peaYoffset-.2) {PEA-type Circuit:};

    \xTip[black] at (\peaXoffset-.5,\peaYoffset+\Yt);
    \draw[quantInfo, ->] (\peaXoffset-.5,\peaYoffset+\Yt)  -- (\peaXoffset-0,\peaYoffset+\Yt);
    \draw[black, very thick] (\peaXoffset-.5,\peaYoffset+\Yt)  -- (\peaXoffset+8.5,\peaYoffset+\Yt);
    \node[circle, minimum size=16pt, inner sep=0pt] at (\peaXoffset+9,\peaYoffset+\Yt) {\large $\ket{\nu_k}$};

    \node[circle, minimum size=16pt, inner sep=0pt] at (\peaXoffset+.4,\peaYoffset-1) {$\ket{0}$};
    \draw[black, very thick]
        (\peaXoffset+.75,\peaYoffset-1) --
        ++(.8, 0) node[rectangle, draw=black, fill=gateColor, inner sep=0pt, minimum size=16pt] (name1) {$H$} -- 
        ++(2.75,0) node[rectangle, draw=black, fill=gateColor] (name2) {$R_z(\theta_R)$} --
        ++(2,0) node[rectangle, draw=black, fill=gateColor] (name3) {$QFT^{-1}$} --
        ++(2,0) node (controlEnd) {};
    \measure at (\peaXoffset+8,\peaYoffset-1);
    \draw[black, very thick]
        (\peaXoffset+.8+2, \peaYoffset-1) node[circle,inner sep=0pt,draw, fill=black, minimum size=.15cm] {.} --
        +(0, \Yt+1) node[rectangle, draw=black, fill=gateColor, inner sep=4pt, minimum size=16pt] (name) {$U^{j}$};

\draw[classInfo, ->] ($(classY)+(delClassYY)+(\peaXoffset-2.25, 0)$)--($(\peaXoffset-2.25, \peaYoffset+\Yt+.75)$) node[pos=.3, classicTextNode]{\tiny instructions};
\xTip[blue] at (\peaXoffset-2.25, \classY+\delClassYY);

\draw[classInfo, ->] (\peaXoffset+4.3,\classY+\delClassYY)--(\peaXoffset+4.3,\peaYoffset-.65) node[pos=.2, classicTextNode]{\tiny $\theta_R$};
\xTip[blue] at (\peaXoffset+4.3,\classY+\delClassYY);

\draw[classInfo, ->] (\peaXoffset+8,\peaYoffset-.7)--(\peaXoffset+8,\classY+\delClassYY) node[pos=.48, classicTextNode]{\tiny $\tilde{C}(\ket{\Phi}, \theta_R)$};
\xTip[blue] at (\peaXoffset+8,\peaYoffset-.7);


\end{tikzpicture}
    \caption{Diagram of variational classical-quantum system. Classical processes are indicated by double blue lines and quantum processes by single black lines. Quantum gates are shown in orange and measurement gates in red. The (potentially high-dimensional) PEA-type circuit is simplified from the typical iterative PEA in that $U$ need not be raised to high orders ($d_c^x$) corresponding to desired eigenphase precision and the measurement gate need only distinguish between the $\ket{0}$-state and the not$(\ket{0})$-state. The (negative) cost function (estimate) $\tilde{\mathcal{C}}$ is returned after a predetermined number of trials of the quantum circuit, approximating a probability.}
    \label{fig:Variational}
\end{figure}

The non-deterministic nature of the IPEA (in the non-eigenstate case) disqualifies the circuit from use as an eigenphase estimator in the standard approach. The SPEA instead considers the probabilistic outputs of the IPEA (and PEA) as valuable information which -- when coupled with a classical controller as in Figure~\ref{fig:Variational} -- allows quantum PEA-like hardware to be used in a variational approach to determine any unknown eigenphase-eigenstate pair. The quantum hardware required is that of a traditional PEA with single-dit precision ($n=1$) and the rotation gate standard to the IPEA (i.e. an iterative PEA with $x$ set to 0). The classical controller determines $\ket{\Phi}$ and $\theta_R$ which are used in the PEA-type circuit. Multiple trials of the quantum circuit are run to approximate the probability $\tilde{\mathcal{C}} \approx \mathcal{C}(\ket{\Phi}, \theta_R)$ (of Equation~\ref{eq:Chat}). Note that the PEA-like circuit need only detect two measurement outcomes: $\ket{0}$ and not$(\ket{0})$, further reducing hardware requirements compared to typical high-dimensional PEAs.
Treating the estimate $-1 \cdot \tilde{\mathcal{C}}(\ket{\Phi}, \theta_R)$ as a cost function in an optimization process (making $\tilde{\mathcal{C}}$ the negative cost function), the classical controller adjusts $\ket{\Phi}$ and $\theta_R$, until the quantum circuit near-deterministically returns $\ket{0}$ as the output state. When $\tilde{\mathcal{C}}(\ket{\Phi}^*, \theta_R^*)\approx 1$, the classical controller has found the (approximate) eigenstate $\ket{\Phi}^*$ and the associated eigenphase $\theta_R^*$.

The quality of the eigenstate $\ket{\Phi}^*$ and eigenphase $\theta_R^*$ retrieval can be quantified by $\mathcal{C}^* = \mathcal{C}(\ket{\Phi}^*, \theta_R^*)$. $\mathcal{C}^*$ can both (1) determine the maximum distance from the eigenphase $\theta_R$ to the nearest eigenphase $\theta_k$ and (2) find the fidelity of $\ket{\Phi}^*$ to actual eigenstate(s). Derivations of both are provided in Appendix~\ref{appendix:CMetric}.

The (negative) cost function $\mathcal{C}$ acts as a metric for quality of eigenvalue-eigenstate retrieval as shown in Appendix~\ref{appendix:CMetric}; by finding $\ket{\Phi}$ and $\theta_R$ which maximize this metric, we arrive at good estimates for an eigenstate ($\ket{\Phi}$) and eigenphase ($\theta_R$) pair. Following is the classical algorithm used to maximize $\mathcal{C}$, which is similar to a gradient search algorithm:
\begin{enumerate}
    \item The classical controller chooses a $\ket{\Phi}$ at random
    \item The classical controller constructs an orthogonal basis $\{\ket{B_m }\}$ including $\ket{\Phi}$
    \item
    \begin{itemize}
        \item (Standard Method: viable when the quantum circuit can measure output bins $\ket{0}$ and not$(\ket{0})$)\\ The quantum circuit evaluates $\tilde{\mathcal{C}}(\ket{\Phi}, \theta_R)$ over a range of $\theta_R$ and returns the maximum value $\mathcal{C}^*$
        \item (Alternative Method: viable when the quantum circuit can measure all $d_c$ output bins: $\ket{0}$ through $\ket{d_c -1}$.)\\ The quantum circuit evaluates $\tilde{\mathcal{C}}(\ket{\Phi}, 0)$ and uses this result to approximate the eigenphase $\theta^*$. The quantum circuit then evaluates $\tilde{\mathcal{C}}(\ket{\Phi}, \theta_R = \theta^*)$ and returns $\mathcal{C}^*$.
        
    \end{itemize}
    \item For all $m=0$ to $2d_t - 1$, we set a=1 and the following occurs:
    \begin{itemize}
        \item if $m\geq d_t$ then $z=\sqrt{-1}$. Otherwise $z=1$.
        \item the classical controller generates the new state:
            \begin{align}
            \begin{split}
                \ket{\Phi'} &= \frac{\ket{A}}{\sqrt{\braket{A|A}}}\\
                    &\text{where } \ket{A} = \ket{\Phi} + z\cdot a \cdot (1-\mathcal{C}^*)\ket{B_{m \mod d_t}}
            \end{split}
            \end{align}
        \item $\ket{\Phi'}$ is fed to the quantum circuit, the maximum value returned is $\mathcal{C}^{'*}$
        \item if $\mathcal{C}^{'*} > \mathcal{C}^{*}$, then $\ket{\Phi} = \ket{\Phi^{'}}$ and $\mathcal{C}^*=\mathcal{C}^{'*}$. Otherwise $\ket{\Phi}$ and $\mathcal{C}^{*}$ are unchanged. 
    \end{itemize}
    \item If $\ket{\Phi}$ was not updated during step 4, set $a=a/2$ and repeat step 4.
    \item If $\mathcal{C}^*$ is greater than the stopping condition or the maximum run-time has been exceeded, the classical controller concludes and returns $\ket{\Phi}$, $\mathcal{C}^*$, and $\theta_R^*$. Otherwise the process continues from step 2.
\end{enumerate}

A few observations on the optimization process may be made. For each iteration, at least $2d_t$ distinct input states are used. For each of these input states a set of $\{ \theta_R \}$ is applied (when using the `standard approach' in step 3). Initially, the $\{ \theta_R \}$ range from $0$ to $1$ with coarse resolution; as the optimization proceeds, $\{ \theta_R \}$ will become fine and include phases from a limited region. Notably, we can choose to run the optimization process limiting $\{ \theta_R \}$ to a narrow range of space from the outset. In this way, we may choose to look only for ground state (small $\theta$), principle (large $\theta$), or any other particular solutions to Equation~\ref{eq:eigen}. In addition, we may eliminate known eigenstates or directions not of interest by excluding them from $\{\ket{B_m }\}$ (step 2) each iteration. In this fashion, the SPEA may be used to determine a complete (or partial) spectral decomposition of $\hat{U}$. Finally, we note while the hardware conventional to a PEA is utilized, this system is superior  to the original PEA, as it determines both the eigenstate and the eigenphase, given no prior knowledge.

\FloatBarrier
\section{Statistical PEA Simulation}
We test the proposed algorithm on the IBM Q platform and on a local computer. In both cases, a classical computer is used to simulate the $\mathcal{C}$ parameter (of Equation~\ref{eq:Chat}) delivered by a quantum circuit. These simulations of a quantum system are ideal: neither the IBM Q nor the local computer simulations include any noise terms. I.e. all quantum gates are assumed to operate with perfect fidelity. The IBM Q trials study the convergence of the optimization algorithm to any single eigenstate on 2- and 4-dimensional systems. The local computer simulations run a \textit{full} spectral decomposition on a 16-dimensional system with various control levels $d_c$.

Both simulations apply the variational algorithm as defined in Section~\ref{Sec:SPEA theory}, with one primary difference: the local computer simulations follows the primary method of step 3 whereas the IBM Q simulations follow the alternative method. The IBM Q simulation runs one measurement with $R_z(\theta_R=0)$ and applies the eigenphase estimation methodology introduced in the Discussion of \cite{lu2020quantum} -- under the (inaccurate first, but increasingly accurate) assumption that the input state is an eigenstate -- to obtain a phase estimate $\theta^*$. Then, the measurement is run with $R_z(\theta_R= -\theta^*)$ to obtain the metric $\mathcal{C}$ used for the optimization. By contrast, the local computer's simulations follow the primary method, picking a representative sample of input phases to apply to the $R_z$ gate and selecting the largest $\mathcal{C}$ that arises. The local computer's simulations therefore require more runs of the quantum circuit per trial, but only require two control-qudit detectors: one for the $\ket{0}$ state and one for the not($\ket{0}$), whereas the IBM Q methodology needs one detector for each control level ($\ket{0}, \ket{1}, ..., \ket{d_c - 1}$). The alternative approach (or some hybrid approach) is generally preferable if the hardware is available for $d_c$ detectors.

\label{Sec:SPEA simulation}
\subsection{Qiskit Simulation}
\label{subsec:Qiskit Simulation}
On the IBM Q experience platform,  we developed our quantum algorithms with Qiskit, the python-based programming package provided by IBM Q which offers all the facilities to design, simulate and execute quantum algorithms on IBM's quantum computers \cite{Qiskit}. In this section we present the simulation results of the SPEA on the Qiskit quantum simulator.

Three sets of simulations are run on the IBM Q, one with 2-dimensional operator $U_1$ and the other two with 4-dimensional operators $U_2$ and $U_3$, the matrix forms of which are shown in Appendix~\ref{appendix:IBMQ}. $U_1$ and $U_2$ are operators directly built with the default gates offered by the IBM Q and $U_3$ is a unitary exponentiation based on the Hamiltonian of the hydrogen molecule generated with  Bravyi-Kitaev transformation\cite{seeley2012bravyi}. The second quantization Hamiltonian of a hydrogen molecule with a bond length $0.74$\r{A}  is calculated by the $STO-3G$ minimal basis using PySCF \cite{PySCF2018} and the transformation is done by OpenFermion\cite{Openfermion2020}. We encode the matrix into the ``Operator" class provided by Qiskit\cite{Qiskit}. In the simulations of each unitary operation $U_i$ where $i=1,2,3$, we start with the input states that are good approximations of one of the operator's eigenstates and then move to input states which are nearly equal-distance from every eigenstate. We quantify the distance of the input state $\ket{\Phi}$ to its nearest eigenstate $\ket{\nu}$ by taking the absolute inner product $\lvert \braket{\Phi | \nu}\rvert$ as reported in Table~\ref{table:IBM results}. In each simulation we run the same input state $20$ times and set the maximum iteration number to be $50$ (to save the resources) and the stopping condition, which is the difference between the $\mathcal{C}$ factor and $1$, to be $10^{-4}$. The stopping condition is set so that when it is met we will have a reasonably good approximation of the eigenstate.  We then calculate the average number of iterations and standard deviation of the number of iterations required to exceed the stopping condition.  Most trials reach the stopping condition before exceeding the iteration limit and give a good approximation of one of the eigenstate-eigenphase pair, as indicated by the low mean phase error reported. The results are shown in Table~\ref{table:IBM results}.

For each operator $U_1, U_2, U_3$, input states which are initially close to an eigenstate (input states with a large absolute inner product) have lower required iteration number than those which are initially far from all eigenstates (low absolute inner product). Appreciate that the eigenstate converged to is non-deterministic, as the optimizer itself is non-deterministic due to randomness added by the random orthogonal basis in step 2. In other words, added randomness may converge the input state to an eigenstate other than the closest eigenstate. The mean phase error recorded in Table~\ref{table:IBM results} is calculated by taking the absolute value of the difference between the eigenphase of the converged input state and the true eigenphase of the eigenstate that the input state converged to. As the input state can converge to different eigenstates in the simulation, we report the absolute phase error rather than the error percentage. No correlation between the phase error and the absolute inner product is apparent, indicating the quality of the final eigenphase-eigenstate pair is agnostic to the proximity of the initial input state to any eigenstate. Variations in mean phase error are likely a function of which particular eigenphase-eigenstate pair was converged to. During the simulation of $U_3$ with an input state of equal weight combination of all the eigenstates -- i.e. the hardest input state to converge to an eigenstate -- there are few cases that the iteration limit is reached and the simulation did not reach the stopping condition. This can usually be fixed by increasing the iteration limit.

In summary, these results indicate that the SPEA method is capable of delivering high-quality estimates of eigenphase-eigenstate pairs with no prior knowledge of the operator's eigenstates, in the case of both arbitrary ($U_1, U_2$) and physically relevant ($U_3$) operators. The quality of the estimates is not influenced by prior system knowledge; however, the resources required to deliver an eigenstate-eigenphase pair may be reduced with prior knowledge.

\begin{table}[]
\begin{center}
\begin{tabular}{||l | l c | ll | l ||}
\hline\hline
\multirow{ 2}{*}{Operation} & \multirow{ 2}{*}{Input State}  & Abs. Inner  &   \multicolumn{2}{c |}{Iteration}    &    \multicolumn{1}{c ||}{Phase Error}\\
    &       &  Product    & Mean  & S.D.  &  \multicolumn{1}{c ||}{Mean}\\ \hline
$U_1$      &  (0.1951, 0.9808)&0.98 &   6.20          &    2.82  & $1.099\cdot 10^{-2}  $                 \\
          & (0.3827,0.9239)&0.92   & 8.15            &  3.41   & $1.005\cdot 10^{-2}  $                     \\
           &  (0.7071,0.7071) &0.71  & 8.90           &  3.34    & $1.005\cdot 10^{-2}   $                    \\ \hline
$U_2$      &  (0 , 0, 0.7432, 0.6690) &0.99       & 5.85        & 8.14  & $2.083\cdot 10^{-2} $          \\
           & (0 , 0,0.6690, 0.7432 ) &0.99       &  6.7          &  10.42&$ 2.168\cdot 10^{-2}$          \\
          &(0,0,1,0)   &0.71& 17.7             &   6.06  & $1.663\cdot 10^{-2} $     \\
           &  (1,0,0,0)  & 0.71&   23.05            &  11.22 & $2.167\cdot 10^{-2} $       \\
                 &   (0.7071, 0   , 0.7071, 0 )&0.50          &  21.3       & 10.71 & $2.262\cdot 10^{-2}$   \\\hline
$U_3$      & (-0.1379,  0 ,  0  ,  0.9904) &0.99   & 1.15         &  0.36   & $1.885\cdot 10^{-2} $         \\
           &  (0    , 0.7807, 0.6247, 0) &0.99          & 1.1              & 0.3  & $1.508\cdot 10^{-2}  $        \\
           &(0,1,0,0)  & 0.71     &  4.35              &  4.17   & $1.414\cdot 10^{-2}  $               \\
          & (0.7071, 0 ,0  , 0.7071 )&0.62 &  4.15        &1.01   & $1.570\cdot 10^{-2} $        \\
           &  (0.5774, 0.5774, 0        , 0.5774)&0.51     &  21.5        &    11.06  & $2.199\cdot 10^{-2}  $  \\ \hline \hline
\end{tabular}
 \caption{
 IBM Q Qiskit QASM simulator results. Three unitary operators are simulated on the IBM Q platform. The SPEA is run 20 times starting from each input state. The distance from the input state to the nearest eigenstate is quantified by the absolute inner product of the two vectors (inner product 1 being identical and smaller values indicting greater difference). The table records the mean and standard deviation (S.D.) of the number of iterations needed to reach the stopping condition ($1-\tilde{\mathcal{C}} = 10^{-4}$). The average absolute phase error is reported in radians. For each $U_i$ the input states range from good approximations of one of the operator's eigenstates to input states which are nearly equal-distance from every eigenstate. See that the iteration mean tends to increase with decreasing inner product but the ultimate phase error is generally agnostic to the input state difference.}
 \label{table:IBM results}
 \end{center}
\end{table}

\subsection{Full Spectral Decomposition}
\label{subsec:Full Spectral Decomposition}
The statistical approach differs from some other variational approaches~\cite{purityVariationalDiag2019} in that it does not diagonalize the input state matrix, but solves for only one eigenphase-eigenstate pair. This allows for significant reduction in the number of parameters (and iterations) needed to perform the optimization. However, as shown below, a complete spectral decomposition is realizable. As a representative case, we consider the $16$-by-$16$ Hamiltonian $\mathcal{H}_{H_2O}$ for the water molecule $H_2O$ with the H-O-H angle at $104.5^{\circ}$ and the bond length at $1.0$ a.u. given in Appendix~\ref{appendix:eigendecomp}~\cite{bian2019quantum}. The Hamiltonian is converted to a unitary exponentiation, $U_{H_2O} = e^{i\mathcal{H}_{H_2O}}$, and the matrix's spectral decomposition is simulated with the statistical variational algorithm (SPEA) on a local computer.

\begin{figure}
\centering
\begin{tikzpicture} 
	\begin{axis}[
		ymin=.9, ymax=1,
		xmin=1.5, xmax=8.5,
		axis y line*=left,
		xlabel={control level $d_c$},
		ylabel={Fidelity},
		title={Eigendecomposition Quality vs $d_c$},
		xmajorgrids,
		ymajorgrids
		]
		\addplot+[only marks, mark=*,
			mark options={scale=1, fill=black},
			style={solid, fill=gray}]
 		coordinates { 
 				(2, 0.98383 )
			(2, 0.96592 )
			(3, 0.98125 )
			(4, 0.9859 )
			(5, 0.98553 )
			(6, 0.98904 )
			(7, 0.99105 )
			(8, 0.99169 )
		};
			\draw [thick,|-| ] (2, 0.97905)--(2, 0.98968);
		\draw [thick,|-| ] (2, 0.95824)--(2, 0.97517);
		\draw [thick,|-| ] (3, 0.97608)--(3, 0.9864);
		\draw [thick,|-| ] (4, 0.98237)--(4, 0.99007);
		\draw [thick,|-| ] (5, 0.98132)--(5, 0.99097);
		\draw [thick,|-| ] (6, 0.98591)--(6, 0.99261);
		\draw [thick,|-| ] (7, 0.99055)--(7, 0.99496);
		\draw [thick,|-| ] (8, 0.98829)--(8, 0.99592);

	\end{axis}

	\begin{axis}[
		xmin=1.5, xmax=8.5,
		ymin=0, ymax=.1,
		ylabel style={align=center},
		ylabel={Eigenphase Error \\ (radian)},
		axis y line*=right,
		axis x line=none,
		ytick={0,.02,.04,.06,.08,.1},
    	yticklabels={$0$,$0.02$,$0.04$,$0.06$,$0.08$,$0.1$},
		axis line style={blue},
		every axis label/.append style ={blue},every tick label/.append style={blue}  
	]
		\addplot+[only marks, mark=square*,
			mark options={scale=1, fill=white!20!blue},
			style={solid, fill=gray}]
 		coordinates { 
 				(2, 0.028423 )
			(2, 0.043381 )
			(3, 0.031195 )
			(4, 0.023999 )
			(5, 0.018641 )
			(6, 0.015261 )
			(7, 0.013699 )
			(8, 0.011976 )
		};
		\draw [blue, thick,|-| ] (2, 0.024071)--(2, 0.031693);
	\draw [blue, thick,|-| ] (2, 0.035915)--(2, 0.048816);
	\draw [blue, thick,|-| ] (3, 0.026884)--(3, 0.034059);
	\draw [blue, thick,|-| ] (4, 0.021583)--(4, 0.025266);
	\draw [blue, thick,|-| ] (5, 0.016604)--(5, 0.020416);
	\draw [blue, thick,|-| ] (6, 0.013709)--(6, 0.016505);
	\draw [blue, thick,|-| ] (8, 0.010531)--(8, 0.013368);
	\draw [blue, thick,|-| ] (7, 0.01179)--(7, 0.015062);
	
	\end{axis}
\end{tikzpicture}
\caption{(Left axis; black; circles) mean fidelity achieved for optimization simulation at control level $d_c$, with error bars from the 25th to 75th percentile. (Right axis; blue; squares) mean phase error (per phase) for optimization simulation at control level $d_c$, with error bars from the 25th to 75th percentile.}
\label{fig:DecompositionResults}
\end{figure}
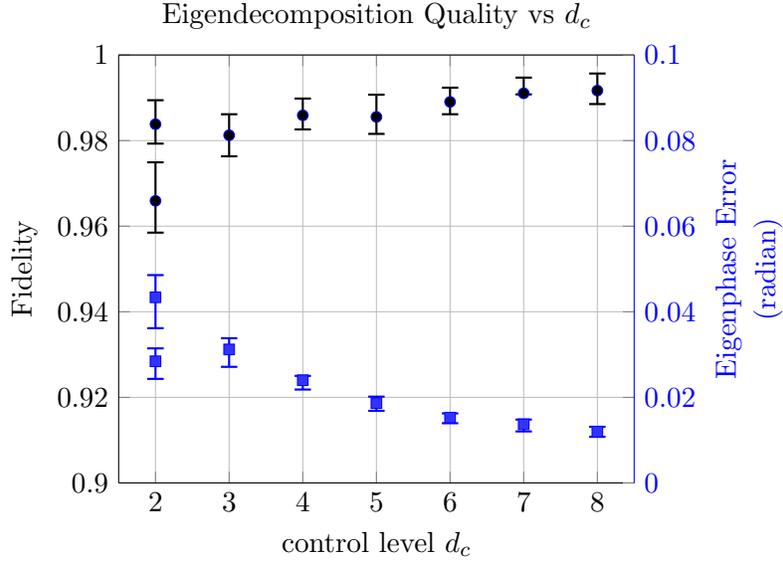

The simulation is run for various control levels $d_c$ until 120 successful spectral decompositions are achieved. Each of the 16 eigenphases are retrieved in a random order. The optimization runs until $\mathcal{C}^* \geq C_{goal}$. If the optimizer is unable to reach $C_{goal}$, the process for that eigenphase will conclude so long as $\mathcal{C}^* \geq C_{req}$. If $ C_{req}$ is not met, the entire spectral decomposition is abandoned and the trial is classified as failed. Generally, $C_{goal}$ is achieved for the first 10 eigenvalues and the latter 2 to 6 eigenvalues must settle at a lower value (due to small cumulative errors). Results are recorded in Table~\ref{table:DecompositionResults} and plotted in Figure~\ref{fig:DecompositionResults}.
\begin{table}
	\begin{center}
	\begin{tabular}{|| c | c c | c c || c | c || c |c ||}
		\hline\hline
		\multirow{ 2}{*}{$d_c$} & \multirow{ 2}{*}{$C_{goal}$} & \multirow{ 2}{*}{$C_{req}$} & \multirow{ 2}{*}{Trials}	& \multirow{ 2}{*}{Fails}	& \multicolumn{2}{c ||}{Fidelity} & \multicolumn{2}{c ||}{Phase Error} \\
 			&	&	& 	&	&	Mean &  S.D. & Mean &S.D.\\ \hline\hline	
 		2 & 0.999 &  0.95  &  120 &  77 &   0.984 &    $7.29\cdot 10^{-3}$ & $2.84\cdot 10^{-2}$ & $10.7\cdot 10^{-2}$\\	
 		2 & 0.995 &  0.9  &  120 &  13 &    0.966 &    $10.68\cdot 10^{-3}$ & $4.34\cdot 10^{-2}$ & $17.3\cdot 10^{-2}$\\
 		3 & 0.995 &  0.9  &  120 &  17 &    0.981 &    $6.62\cdot 10^{-3}$ & $3.12\cdot 10^{-2}$ & $9.14\cdot 10^{-2}$\\
 		4 & 0.995 &  0.9  &  120 &  32 &    0.986 &    $5.32\cdot 10^{-3}$ & $2.40\cdot 10^{-2}$ & $7.36\cdot 10^{-2}$\\
 		5 & 0.995 &  0.9  &  120 &  30 &    0.986 &    $7.10\cdot 10^{-3}$ & $1.86\cdot 10^{-2}$ & $4.98\cdot 10^{-2}$\\
 		6 & 0.995 &  0.9  &  120 &  26 &    0.989 &    $4.80\cdot 10^{-3}$ & $1.53\cdot 10^{-2}$ & $3.80\cdot 10^{-2}$\\
 		7 & 0.995 &  0.9  &  120 &  60 &    0.991 &    $7.05\cdot 10^{-3}$ & $1.37\cdot 10^{-2}$ & $3.65\cdot 10^{-2}$\\
 		8 & 0.995 &  0.9  &  120 &  132 &   0.992 &    $6.13\cdot 10^{-3}$ & $1.20\cdot 10^{-2}$ & $3.54\cdot 10^{-2}$\\
 		\hline\hline
	\end{tabular}
	\end{center}
	\caption{Statistics for 120 successful trials of complete spectral decomposition of $U_{H_2O}$ matrix. $C_{goal}$ is the $\tilde{C}$ value the optimizer attempts to reach, and generally does reach for at least the first 10 (of 16) eigen-estimates. $C_{req}$ is the $\tilde{C}$ value the optimizer is required to reach for all eigen-estimates, else the trial is abandoned. For successful trials, the  mean Fidelity and standard deviation are reported, as well as the mean eigenphase error (in radians). Note that $d_c=2$ was run on two different $C_{goal}$, $C_{req}$ levels for comparison.}
\label{table:DecompositionResults}
\end{table}

To determine the fidelity of the spectral decomposition, the retrieved eigenphase-eigenstate pairs, $(\theta_k, \ket{\nu_k})$ were used to create the matrix,
\begin{align}
    U_{retrieved} = \sum_k e^{i2\pi\theta_k} \ket{\nu_k} \bra{\nu_k}.
\end{align}
Letting $M= U_{H_2O}^\dagger U_{retrieved}$, the fidelity is defined as
\begin{align}
    \text{fidelity} = \frac{1}{n\cdot (n+1)} \big( Tr(MM^\dagger ) + \abs{Tr(M)}^2\big)
\end{align}
following the average fidelity definition of~\cite{pedersen2007fidelity} where $n$ is matrix dimension (i.e. $n = 16$). The reported phase error is the average absolute phase error over all 16 phases,
\begin{align}
    \text{phase error} = \frac{\sum_k \abs{2\pi\theta_k -2\pi\theta_{k, true}}}{n}.
\end{align}

Note that the $d_c=2$ was run for two different sets of $C_{goal}$ and $C_{req}$. Increasing these values increased the optimization failure rate, but also improved decomposition fidelity and reduced the average eigenphase error. The high failure rate suggests superior results will be achieved by increasing $d_c$, the number of control levels, over increasing $\mathcal{C}^*$ (analogous to $C_{goal}$), when possible. This is expected, as increasing $d_c$ leads to a narrower cost function. Overall, these results indicate both the viability of the SPEA for full and partial eigenphase recovery and provides an example of a quantum algorithm which benefits from working with high-dimensional quantum states, i.e. qudits.

\FloatBarrier
\section{Conclusion}
\label{Sec:conclusion}
In this work, we have proposed  a  novel  statistical variational  approach (SPEA) to  the quantum phase  estimation  algorithm (PEA). From the probabilistic output of a PEA circuit using non-eigen input states, we have defined a statistical metric $\mathcal{C}$ indicating the proximity  of any given input state to the nearest eigenstate and develop an optimization process that can variationally retrieve all the eigenstate-eigenphase pairs of a given unitary operator. The SPEA takes advantage of the hardware intended for the Iterative PEA and therefore requires no novel quantum hardware development. The main disadvantage of the SPEA is the non-deterministic nature of the measurements requires running the quantum circuit repeatedly for each measurement setting. However, in the near-term era, repeated runs of a quantum circuit per measurement is already the norm, due to noise and imperfect gate fidelity. One of the main advantages of the SPEA compared to the PEA and IPEA is the ability to systematically find both the eigenstates and associated eigenphases, rather than just the eigenphases.

The simulations on the IBM Q platform with Qiskit proves the feasibility of applying the SPEA on standard quantum computation platforms. On the local computer, the full spectral decomposition of the operator generated from the water molecule Hamiltonian demonstrates the viability of the SPEA for applications in quantum chemistry. The ability to retrieve eigenstates and efficiency (in terms of low iterations requirement) of this method shows the improvement to the original PEA methods and offers the clear potential to work with larger physical and chemical systems.

Future work includes improving the optimization process with a more sophisticated algorithm for the classical controller. In addition to improving efficiency and failure rate, this may also improve the accuracy of the eigenphase-eigenstate retrieval as well as the fidelity of the full spectral decomposition.  The efficiency and the viability of our methods enable us to simulate more complex systems in quantum chemistry. Future work also includes implementing this method on real computational systems provided by the IBM Q and also on a photonic platform with high-dimensional control qudit capabilities.

\section*{Acknowledgements}
We would like to acknowledge the financial support by the National Science Foundation under award number 1839191-ECCS
\printbibliography

\newpage
\appendix
\section{Quality of Eigenphase-Eigenstate retrieval}
\label{appendix:CMetric}

\textbf{Preliminary note:} All phases $\theta\in [0, 1)$. The difference between two phases can be found via the function 
\begin{align}
    d(\theta_a, \theta_b) \equiv  \frac{\text{angle}(e^{i2\pi (\theta_a - \theta_b)})}{2\pi}.
\end{align}
I.e. $d(\theta_a, \theta_b) \in [-.5, .5)$. In short, as the phase wraps around the phase difference also wraps around. E.g. $d(\frac{1}{8}, \frac{3}{4}) = \frac{3}{8}$, not $\frac{-5}{8}$. And $d(\frac{3}{4}, \frac{1}{8}) = \frac{-3}{8}$, not $\frac{5}{8}$. Like usual differences, $d(\theta_a, \theta_b) = -d(\theta_b, \theta_a)$. Appreciate that -- for $\mathcal{P}_0$ of Equation~\ref{Eq:Prob0} -- $\mathcal{P}_0 (d(\theta_a, \theta_b)) = \mathcal{P}_0 (\theta_a - \theta_b) \forall \theta_a, \theta_b$. 

Here, we quantify the quality of the eigenstate $\ket{\Phi}^*$ and eigenphase $\theta_R^*$ using $\mathcal{C}^* = \mathcal{C}(\ket{\Phi}^*, \theta_R^*)$. When $\mathcal{C}^*$ is greater than the largest (non-global) local maximum of $\mathcal{P}_{\theta}$ (of Equation~\ref{Eq:Prob0}), then $\theta_R^*$ must be within the primary lobe of $\mathcal{P}_{\theta}$ (examples shown in Figure~\ref{fig:Pcurves}). That is, when
\begin{align}\label{eq:lobeCondition}
    \mathcal{C}^* > \max_{\zeta \in [\frac{1}{d_c}, \text{ } 0.5]} \mathcal{P}_{0}(\zeta)
\end{align}
we are within the primary lobe of $\mathcal{P}_{\theta}$. Let $\theta_{k^*}$ be the eigenvalue closest to $\theta_R$ and define $\delta^* \equiv d(\theta_{k^*}, \theta_R$). When Equation~\ref{eq:lobeCondition} is true, then $\mathcal{P}_{0}(\delta^*) \geq \mathcal{P}_{0}(\theta_k - \theta_R) \forall k$ and 
\begin{align}\begin{split}
\mathcal{C}^* &= \sum_{k=0}^{d_t - 1} \abs{\braket{\theta_k | \Phi}}^2 \mathcal{P}_{0}(\theta_k - \theta_R)     \leq \sum_{k=0}^{d_t - 1} \abs{\braket{\theta_k | \Phi}}^2 \mathcal{P}_{0}(\delta^*) \\
    &= \mathcal{P}_{0}(\delta^*) \sum_{k=0}^{d_t - 1} \abs{\braket{\theta_k | \Phi}}^2 = \mathcal{P}_{0}(\delta^*) \\
\mathcal{C}^* &\leq  \mathcal{P}_{0}(\delta^*)
\end{split}\end{align}

As $\mathcal{P}_{0}$ is symmetric and monotonic within the primary lobe,
\begin{align}
    \abs{\delta^*} \leq \mathcal{P}_{0}^{-1}(\mathcal{C}^*).
\end{align}
Therefore the estimated eigenphase $\theta_R^*$ is within $\pm \mathcal{P}_{0}^{-1}(\mathcal{C}^*)$ of the nearest eigenphase (whenever Equation~\ref{eq:lobeCondition} is met).

Now to quantify the eigenstate estimate. Define a $\Delta$-eigenstate $\ket{\nu_\Delta}$ as any superposition of eigenstates where the corresponding eigenphases are within $\pm \Delta$ of $\theta_R^*$ 
\begin{align} \label{eq:degenEigenstate}
\begin{split}
    \ket{\nu_\Delta} &= \sum_m \alpha_m \ket{\nu_m} \\
    &\text{where } \lvert d(\theta_m, \theta_R^*) \rvert \leq \Delta \forall m 
\end{split}
\end{align}
(and where $\sum_m \abs{\alpha_m}^2 = 1$). That is, $\ket{\nu_\Delta}$ is a superposition of eigenstates (indexed $\{ m \}_\Delta$) that are \textit{nearly} degenerate: the corresponding eigenphases are all within $2\Delta$ of one another. Proceeding from Equation~\ref{eq:Chat} (whenever Equation~\ref{eq:lobeCondition} holds),
\begin{align}\label{eq:fidelitySt}
\begin{split}
\mathcal{C}^* = \mathcal{C}(\ket{\Phi}^*, \theta_R^*) 
    &= \sum_{k=0}^{d_t - 1} \abs{\braket{\nu_k | \Phi}}^2 \mathcal{P}_{0}(\theta_k-\theta_R^*) \\
    &\leq \sum_{k \in \{m\}_{\Delta}} \abs{\braket{\nu_k | \Phi}}^2 (1) + \sum_{k\notin \{m\}_\Delta} \abs{\braket{\nu_k | \Phi}}^2\mathcal{P}_{0}(\theta_k - \theta_R^*)\\
    &\leq \sum_{k \in \{m\}_{\Delta}} \abs{\braket{\nu_k | \Phi}}^2 + \sum_{k\notin \{m\}_\Delta} \abs{\braket{\nu_k | \Phi}}^2 \mathcal{P}_{0}(\Delta).
\end{split}
\end{align}
Letting $\sum_{k \in \{m\}_{\Delta}} \abs{\braket{\nu_k | \Phi}}^2 = f$,
\begin{align} \label{eq:fidelityEd}
\begin{split}
\mathcal{C}^*    &\leq f + (1-f) \mathcal{P}_{0}(\Delta)\\
\therefore \text{ }& f \geq \frac{\mathcal{C}^* - \mathcal{P}_{0}(\Delta )}{1 - \mathcal{P}_{0}(\Delta )} 
\end{split}
\end{align}
The estimated eigenstate $\ket{\Phi}^*$ matches some $\Delta$-eigenstate (as defined by Equation~\ref{eq:degenEigenstate}) with fidelity $f$ given by Equation~\ref{eq:fidelityEd} (whenever Equation~\ref{eq:lobeCondition} holds and $\abs{\Delta}  \in [0, \frac{1}{d_c}]$).

\section{Details for the IBM Q SPEA calculations}
\label{appendix:IBMQ}
On IBM Q we realize an SPEA with a four-dimensional control register by using two qubits (the top two rails) as controls. The target is either two- or four-dimensional, using the bottom one or two rails, respectively.

We list out the three operators in matrix form used in our simulations on the IBM Q accompanied by the eigenstates of each matrix as well as showing how we achieved these matrices with the Qiskit.

\begin{figure}
\begin{center}
\includegraphics[width=0.7\linewidth]{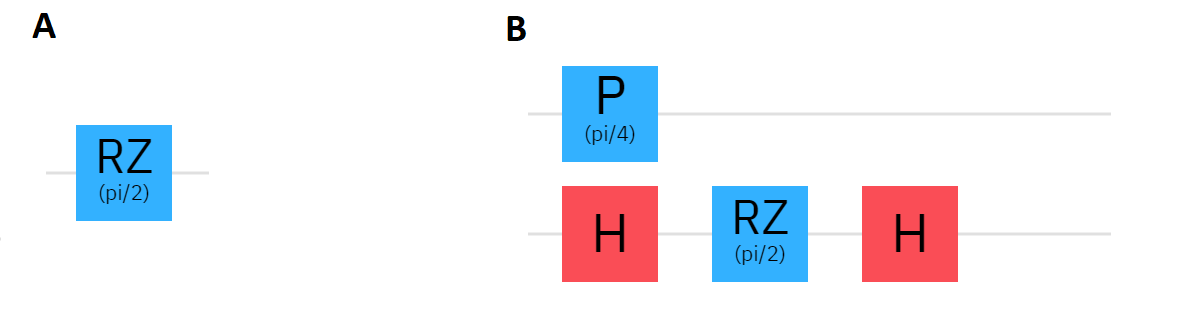}\\
\caption{Gate representation of the unitary operators applied in the simulation. \textbf{(A)} represents operator $U_1$, a 2-dimensional operator applied to a single qubit.  \textbf{(B)} represents operator $U_2$, a 4-dimensional operator applied to two qubits. $RZ$ is the rotation-$Z$ gate, $P$ is the phase gate and $H$ is the Hadamard gate. When a control qubit is present the $RZ$ gate and the $P$ gate become controlled gates. The Hadamard gates will serve as their own inverse and therefore do not need to be implemented as controlled gates.} 
\label{fig:Uni_operators}
\end{center}
\end{figure}

In the following we use the rotation-$Z$ gate as defined by Qiskit~\cite{Qiskit}:
\begin{align}
    RZ(\theta) = \begin{pmatrix}
        e^{-i \theta / 2}   &   0\\
        0                   & e^{i \theta / 2}
    \end{pmatrix}
\end{align}
as well as the phase gate:
\begin{align}
    P(\theta) = \begin{pmatrix}
        1   &   0\\
        0   & e^{i\theta}
    \end{pmatrix}
\end{align}
and the Hadamard gate:
\begin{align}
    H = \frac{1}{\sqrt{2}}\begin{pmatrix}
        1   &   1\\
       1   & -1
    \end{pmatrix}
\end{align}
The first operator $U_1$ is a single qubit rotation-$Z$ gate with $\theta = \pi/2$,
\begin{equation}
    U_1 = RZ(\pi/2) = \begin{pmatrix}
    e^{-i\pi / 4} & 0 \\
    0& e^{i\pi / 4} 
    \end{pmatrix} \; \text{with eigenstates}\: 
    v_1= \begin{pmatrix}
    0  \\
    1 
    \end{pmatrix},
    v_2= \begin{pmatrix}
    1  \\
    0 
    \end{pmatrix}
\end{equation}
The second operator is a two qubit operation achieved by a phase gate $P(\theta = \pi/4)$ acting on the first qubit and a rotation-$Z$ gate $RZ(\theta = \pi/2)$ sandwiched between two Hadamard gates $H$, acting on the second qubit. The matrix form is
\begin{align}
    U_2   = P(\pi/4)\otimes \big( H \cdot RZ(\pi/2) \cdot H\big) = \frac{1}{\sqrt{2}}\begin{pmatrix}
      1 &   -i  &   0   &   0\\
      -i    &   1   &   0   &   0\\
      0     &   0   &   e^{i\frac{\pi}{4}}    &   e^{-i\frac{\pi}{4}}\\
      0     &   0   &    e^{-i\frac{\pi}{4}}    &   e^{i\frac{\pi}{4}}
    \end{pmatrix}
\end{align}
with the eigenstates
\begin{equation}
    v_3= \begin{pmatrix}
    0  \\
    0 \\
    1\\
    1 
    \end{pmatrix}
    v_4= \begin{pmatrix}
    0  \\
    0 \\
    1\\
    -1 
    \end{pmatrix},
    v_5= \begin{pmatrix}
    1  \\
    1 \\
    0\\
    0 
    \end{pmatrix},
    v_6= \begin{pmatrix}
    1  \\
    -1 \\
    0\\
    0 
    \end{pmatrix}
\end{equation}
The gate representations of the two operators can be found in Figure \ref{fig:Uni_operators}.

For the third operators we start with the lowest energy Hamiltonian of the hydrogen molecule generated with  Bravyi-Kitaev transformation in $STO-3g$ basis,
\begin{equation}
    H=
\begin{pmatrix}
0.48704885&  0       &  0       &0.18065279 \\
0 & -0.33769999&  0.18065279&0    \\
0.     &  0.18065279& -0.33769999&0 \\
0.18065279&  0&  0&-1.11719411
\end{pmatrix}.
\end{equation}
Then we take the exponential of the Hamiltonian to generate our unitary operators
\begin{equation}
U_3 = e^{i H}=
\begin{pmatrix}
0.8686+0.4687i& 0         &0         & 0.0499 +0.1531i \\
0         & 0.9282-0.3259i &0.0595+0.1695i& 0          \\
0         & 0.0595+0.1695i&0.9282-0.3259i & 0          \\
0.0499 +0.1531i& 0         &0         & 0.4256 -0.8905i
\end{pmatrix}.
\end{equation}
The exponential of the Hamiltonian is done with the scipy.linalg.expm in the scipy python package~\cite{2020SciPy-NMeth}.
The eigenstates are 
\begin{equation}
    v_7= \begin{pmatrix}
-0.1105  \\
    0 \\
    0\\
    0.9939 
    \end{pmatrix}
    v_8= \begin{pmatrix}
    0  \\
    0.7071 \\
   -0.7071\\
   0
    \end{pmatrix},
    v_9= \begin{pmatrix}
    0  \\
    0.7071 \\
    0.7071\\
    0 
    \end{pmatrix},
    v_{10}= \begin{pmatrix}
     -0.9939  \\
   0 \\
    0\\
    0.1105 
    \end{pmatrix}
\end{equation}

\begin{figure}
\begin{center}
\includegraphics[width=0.95\linewidth]{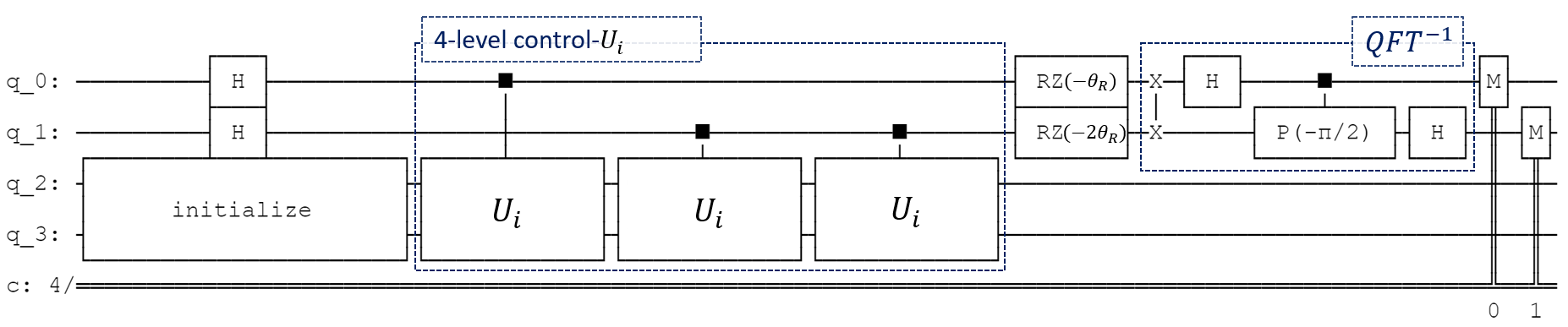}\\
\caption{Illustration of the 4-level control ($d_c=4$) SPEA circuit implemented using Qiskit. $H$ is the Hadamard gate, $RZ$ is the rotation-$Z$ gate, $P$ is the phase gate, and $M$ represents a measurement. The top two rails are the control qubits and realize a 4-dimensional control register. The bottom two rails represents the target register. The initialize block prepares the input state differently throughout the optimization process. $U_i$ represents the unitary the SPEA is running. For $i=2, 3$ we implement the circuit as diagrammed. For $i=1$, the unitary is 2-dimensional and we use only a single target rail. The three control gates together realize a multi(4)-level control gate as described in the main text. The two $RZ$-gates together realize a realize four-level $R_z(-\theta_R )$ as described in the main text. The double line at the bottom represents the classical information retrieved from the measurement gates (a total of 2 bits).} 
\label{fig:QiskitPEA}
\end{center}
\end{figure}

We design the phase estimation algorithms that work with up to two qubits in the control register and four qubits in total as shown in Figure \ref{fig:QiskitPEA}. Due to the restrictions in the qubit numbers, the simulations on the IBM Q are focused on the lower dimensional systems, i.e. one or two qubits in the target register. The rotation $RZ$ gates are applied to each qubit in the control register after the controlled-$U_i$ operations but before inverse Fourier transform. Every time the quantum algorithm is called to generate a new $\mathcal{C}$ factor (following the ``alternative method'' in step 3 of Section~\ref{Sec:SPEA theory}), the algorithm runs twice: the first time the $RZ$ gates are set to zero and the phase factor $\theta_R$ is calculated statistically; the second time the upper $RZ$ gate applies phase $-\theta_R$ and the lower gate applies phase $-2\theta_R$, together acting as a $-\theta_R$-rotation would on a $d_c=4$ qudit system. The classical optimization process described in Section~\ref{Sec:SPEA theory} is implemented using python. The basis set $\{\ket{B_m }\}$  is generated using the Gram-Schmidt methods with the input vector plus a set of linearly independent vectors obtained from a randomly generated unitary matrix. As a deviation from how the algorithm is described in the main text, the search step $a$ factor is set to $1/2$ in step 4 and is doubled in step 5 (rather than halved) if $\mathcal{C}$-factor is not updated, up to seven times. The optimization concludes when the $\mathcal{C}$ factor meets the stopping condition which means the input state is converged to an eigenstate, or when the maximum iteration time is exceeded.

\section{Full $H_2O$ Matrix}
\label{appendix:eigendecomp}
The Hamiltonian of the water molecule with the H-O-H angle at $104.5^{\circ}$ and the bond length at $1.0$ a.u. is calculated by $STO-3G$ minimal basis using PySCF \cite{PySCF2018} and chemistry package provided by the Qiskit\cite{Qiskit}. The 16-by-16 Hamiltonian of the water molecule used for the local computer's spectral decomposition simulations is as follows~\cite{bian2019quantum}. The exponential of the Hamiltonian is done with the MATLAB expm funtion.
\\
{\small
\rotatebox[origin=bl]{90}{
\parbox{\textheight}{
\begin{align}\begin{split}
    &\mathcal{H}_{H_2O} =\\
    &\begin{pmatrix}
0    &    0    &    0    &    0    &    0    &    0    &    0    &    0    &    0    &    0    &    0    &    0    &    0    &    0    &    0    &    0\\ 
0    &    -2.594    &    0    &    0    &    0    &    0    &    0    &    0    &    0    &    0    &    0    &    0    &    0    &    0    &    0    &    0\\ 
0    &    0    &    -2.654    &    0    &    0    &    0    &    0    &    0    &    0    &    0    &    0    &    0    &    0    &    0    &    0    &    0\\ 
0    &    0    &    0    &    -4.583    &    0    &    0    &    0    &    0    &    0    &    0    &    0    &    0    &    0    &    0    &    0    &    0\\ 
0    &    0    &    0    &    0    &    -2.594    &    0    &    0    &    0    &    0    &    0    &    0    &    0    &    0    &    0    &    0    &    0\\ 
0    &    0    &    0    &    0    &    0    &    -4.427    &    0    &    0    &    0    &    0    &    0.054    &    0    &    0    &    0    &    0    &    0\\ 
0    &    0    &    0    &    0    &    0    &    0    &    -4.529    &    0    &    0    &    0.054    &    0    &    0    &    0    &    0    &    0    &    0\\ 
0    &    0    &    0    &    0    &    0    &    0    &    0    &    -5.696    &    0    &    0    &    0    &    0    &    0    &    0    &    0    &    0\\ 
0    &    0    &    0    &    0    &    0    &    0    &    0    &    0    &    -2.654    &    0    &    0    &    0    &    0    &    0    &    0    &    0\\ 
0    &    0    &    0    &    0    &    0    &    0    &    0.054    &    0    &    0    &    -4.529    &    0    &    0    &    0    &    0    &    0    &    0\\ 
0    &    0    &    0    &    0    &    0    &    0.054    &    0    &    0    &    0    &    0    &    -4.428    &    0    &    0    &    0    &    0    &    0\\ 
0    &    0    &    0    &    0    &    0    &    0    &    0    &    0    &    0    &    0    &    0    &    -5.637    &    0    &    0    &    0    &    0\\ 
0    &    0    &    0    &    0    &    0    &    0    &    0    &    0    &    0    &    0    &    0    &    0    &    -4.583    &    0    &    0    &    0\\ 
0    &    0    &    0    &    0    &    0    &    0    &    0    &    0    &    0    &    0    &    0    &    0    &    0    &    -5.696    &    0    &    0\\ 
0    &    0    &    0    &    0    &    0    &    0    &    0    &    0    &    0    &    0    &    0    &    0    &    0    &    0    &    -5.637    &    0\\ 
0    &    0    &    0    &    0    &    0    &    0    &    0    &    0    &    0    &    0    &    0    &    0    &    0    &    0    &    0    &    -6.085\\ 
    \end{pmatrix}
\end{split}
\end{align}
}}}

\end{document}